# Highly efficient light management for perovskite solar cells

Dong-Lin Wang[1], Hui-Juan Cui[1], Guo-Jiao Hou[2], Zhen-Gang Zhu[2,1], Qing-Bo Yan[3] & Gang Su[1]*

[1] School of Physics, University of Chinese Academy of Sciences, P. O. Box 4588, Beijing 100049, China.

[2] School of Electronic, Electrical and Communication Engineering, University of Chinese Academy of Sciences, Beijing 100049, China

[3] College of Materials Science and Opto-Electronic Technology, University of Chinese Academy of Sciences, Beijing 100049, China

*Correspondence and requests for materials should be addressed to G.S. (email: gsu@ucas.ac.cn).

Organic-inorganic halide perovskite solar cells have enormous potential to impact the existing photovoltaic industry. As realizing a higher conversion efficiency of the solar cell is still the most crucial task, a great number of schemes were proposed to minimize the carrier loss by optimizing the electrical properties of the perovskite solar cells. Here, we focus on another significant aspect that is to minimize the light loss by optimizing the light management to gain a high efficiency for perovskite solar cells. In our scheme, the slotted and inverted prism structured $SiO_2$ layers are adopted to trap more light into the solar cells, and a better transparent conducting oxide layer is employed to reduce the parasitic absorption. For such an implementation, the efficiency and the serviceable angle of the perovskite solar cell can be promoted impressively. This proposal would shed new light on developing the high-performance perovskite solar cells.

Photovoltaic (PV) device with high conversion efficiency and low cost are expected for an extensive utilization of solar energy. Recently, the emergence of organic-inorganic halide perovskite materials ($CH_3NH_3PbX_3$, X=Cl, Br, I) opens up new possibilities for cost-effective PV modules [1-4]. In a few short years, the efficiency of perovskite solar cell has skyrocketed from 3.8% to around 20% [5-11]. Many strategies are employed to promote the efficiency of the perovskite solar cells, such as, the interface materials engineering [7,12-14], fabrication processing optimization [6,15-18], with or without mesoporous scaffold design [19-22], and so on. Those schemes mainly focus on improving the electrical properties of the solar cells to minimize the carrier loss attempting to achieve a high conversion efficiency. However, an efficient light management is also significant to enhance the efficiency of the solar cells by trapping more light into the active layers to reduce the light loss. To get high-performance perovskite solar cells, it is quite essential to balance both the electrical and optical benefits of the cells.

In a simple perovskite solar cell, the active layer ($CH_3NH_3PbI_3$) is sandwiched between the hole and electron transport layer (HTL and ETL) [6,12,14,23]. In such a structure, two electrical benefits, a high collection efficiency and a low recombination of carriers, are indispensable to realize a high conversion efficiency. Thus, it is necessary to enhance the material quality of the perovskite to increase the mobility and life times of carriers, and decrease the defect density. Aside from the material quality, decreasing the thickness of the active layer is also a way to implement the above mentioned electrical benefits [24]. Nonetheless, such a thin absorber cannot maintain a high light absorption to excite adequate carries. Light trapping can provide a perfect solution to absorb more light in the thin active layer, ultimately, to realize mutual benefits for both optical and electrical properties of the perovskite solar cells.

A typical perovskite solar cell is shown in Fig 1a, where 80nm thick ITO (indium doped tin oxide) is deposited on a flat glass, followed by 15nm thick PEDOT:PSS (poly(3,4-ethylenedioxythiophene):poly (styrene sulfonate)), 5nm thick PCDTBT (poly(N-9'-heptadecanyl-2,7-carbazole-*alt*-5,5-(4',7'-di(thien-2-yl)-2',1', 3'-benzothiadiazole))), 350nm thick $CH_3NH_3PbI_3$, 10nm thick PC60BM (((6,6)-phenyl-C61-butyric acid methyl ester) and 100nm thick Ag layer. In this architecture, PEDOT:PSS and PC60BM are considered as HTL and ETL to extract the excited carriers in the $CH_3NH_3PbI_3$ layer. Regardless of the carrier loss in the process of transport, the output current is determined by how much light can be absorbed by the active layer. As shown in Fig 1b and 1c, the absorption and reflection efficiency of each section can be analyzed by employing an optical simulation (the optical constants of each layers are measured by Q. Lin *et al*. [14]). The results indicate that the active layer absorbs only 65% of the incident light that can be effectively utilized to excite carriers. The light loss of 2% is in the HTL, ETL and Ag layers, 14% is absorbed by the ITO layer, 4% is reflected by the surface of the glass and 15% escapes from the solar cell. The first and third losses are hard to suppress. However, the parasitic absorption in ITO layer and unabsorbed light can be suppressed by an efficient light management. It is worth noting that to excite carriers there are almost 30% of light that may be reused to excite carriers. To utilize those wasted light, we propose a highly efficient light management scheme to minimize the light loss in the perovskite solar cell, which mainly consists of the design of a high-efficient light trapping structure to suppress light reflection and the use of a better ITO layer to reduce the parasitic absorption.

**Results**

**Design of the high-efficient light trapping structure for perovskite solar cell.** Recently, nano-scaled dielectric and metallic structures based light trapping has been exposed to exhibit excellent capacities to promote the efficiency of the silicon thin-film solar cells [25-28]. Nevertheless, the light trapping is hard to collect all of wavelengths of the utilized light because of the mechanism of the wave optics. It appears that the ray optical-based light trapping can avoid this issue. What is more, the ray optics-based front retroreflector, back reflector and structured substrate are proved to be beneficial to enhance the efficiency of the organic solar cells [29-33]. For perovskite solar cells, the light trapping structures based on the ray optics may be more appropriate to trap broadband sun light and keep low cost benefits. In this work, we propose a cheap light trapping structure based on the prism structured $SiO_2$ for perovskite solar cells. As shown in Fig 2a, the prism structured $SiO_2$ structures are periodically arranged on the top surface of the solar cells. The cross-section of the prism is an equilateral triangular geometry with bottom width L and base angle θ. In the calculation, the size of the width is set large enough (L=10μm) to minimize the impact of light interference.

In a perovskite solar cell, most of light can be absorbed by the active layer in a single light path when the wavelength (λ) of the light is short (λ<500nm). In this situation, the main light loss is the reflection at the front surface of the cell. For the long wavelength (λ>500nm), the unabsorbed light will be reflected back by the bottom Ag layer. So the reuse of the reflected light is a feasible way to reduce the light loss. For this purpose, a proper designed prism structure is capable of controling the light travel path and eliminating the reflected light. Our calculation indicates that the prism with the base angle θ=42 °is the best for light trapping. As shown in Fig 2b, the refraction and the total internal reflection from the prism side can adjust the light direction to realize at least three times light injection when the light is incident from Ⅰ region. Regardless of the light loss at the interface between air and the prism, the total reflection efficiency (R) of the solar cell can be reduced to $R^N$, where N is the number of times that the incident light is injected into the active layer. In this case, three times injection of the incident light is enough to absorb all of the incident light. As a result, the prism based light trapping structures can enhance the absorption efficiency for the active layer, and reduce the total light reflection with wavelength from 300nm to 800nm (as shown in Fig 2c). Note that the above optical benefits are only provided by the incident light from Ⅰ region. The light trapping by the prism can further be improved by a proper utilization of the incident light from Ⅱ region.

Based on the above calculation, we find that the averaged total reflection of the perovskite solar cell without using light trapping structures is below 20%. It is amazing that double reinjection of the incident light by light trapping structure can suppress the reflection to around 4% (R=20%, $R^2$=4%), which is sufficient to trap most of incident light into the solar cell. To maximize the light trapping capacity of the prism structure, it is necessary to reuse the reflected light from both Ⅰ and Ⅱ incident regions. Here, we adopt the slotted prism structures with equilateral triangular cross-section, as shown in Fig 3a, which are located under the prism bottom. The structure of the perovskite solar cell is conformed to the slotted structures whose size should be properly designed to assure the reinjection of the reflected light (details can be found in Supplementary Note1).

Fig 3a shows that the optimized prism with slotted structure can realize at least double light injection for all incident light. The total reflection of the solar cell can be compressed to be under 5% for the light with wavelength from 300nm to 750nm (as shown in Fig 3b). Thus, the light absorption in the active layer is greatly improved by the above strategy. For the perovskite solar cell without light trapping structures, $J_G$ can reach 18.49 mA/cm$^2$ when light is of normal incidence, while with slotted prism based light trapping structure $J_G$ can be increased up to 21.46 mA/cm$^2$. However, the light trapping ability of the slotted prism is sensitive to the incident angle of the light, and under oblique incidence of light, the capacity to reuse the reflected light may be degraded. As shown in Fig 3c, $J_G$ becomes smaller as the oblique angle increases, where $J_G$ larger than 20 mA/cm$^2$ can be sustained only when the oblique angle is less than 30°. Obviously, this small angle cannot satisfy the requirement of sufficient utilization of sun light for a solar cell without a tracking system.

A large serviceable angle of incident light and a high capacity of light trapping are preferable for a better light trapping scheme. Here, the SiO$_2$ inverted prism structure is employed for the perovskite solar cell to achieve the above two purposes. As shown in Fig 4a, the structure of the perovskite solar cell is conformed to the side of the inverted prism, and sun light is incident from the bottom of the prism. In this structure, the unabsorbed light after the first injection can be reused by properly utilizing the total internal reflection of the prism bottom when the light is obliquely incident (see Fig 4b). However, the double injection of the incident light into the active layer can be sustained only when the light is within a serviceable angle determined by the base angle of the prism. This angle is defined by $\alpha<\arcsin(n_{glass} \sin(2\theta-\theta_c))$ for $\theta \leq \theta_c$ and $\alpha=90°$ for $\theta>\theta_c$, where $n_{glass}$ is the reflective index of the glass and $\theta_c \approx 41.5°$ is the critical angle for total reflection of light from glass to air (more details can be found in Supplementary Note2). It can be seen that the high-efficient light tapping can be sustained for all oblique angles by employing the inverted prism with base angle $\theta \geq 42°$.

Now let us look at the optical properties of the perovskite solar cell with the inverted prism (base angle $\theta=42°$) light trapping structure. In this strategy, the double injection caused by the prism of the incident light can promote the average total absorption to approach 95% for broadband wavelengths when the light is in normal incidence (see Fig 4c). The enhancement of the light absorption in the active layer can enhance $J_G$ to around 21.01 mA/cm$^2$. More importantly, this optical benefit can be maintained within a large serviceable angle. As shown in Fig 4d, $J_G$ larger than 20 mA/cm$^2$ can be sustained when the oblique angle is within about 60°. However, the light trapping ability is degraded when the incident angle is larger than this serviceable angle due to the increased reflection at the front surface of the glass and one side of the inverted prism structure. It is interesting to note that this serviceable angle is double that of the perovskite solar cell with slotted prism based light trapping structure.

**Reducing the parasitic absorption in ITO layer.** The designed light trapping structures are capable of suppressing the total reflection to approach the limitation (R≈5%) that only contains the reflection at the front surface of the glass for broadband wavelength. However, a large parasitic absorption in the ITO layer still wastes much of trapped light in proposed perovskite solar cells. On the other hand, improving the material quality of the ITO layer is another viable way to enhance the light absorption in the active layer. The ITO layer studied in this work is based on the commercial ITO patterned glass electrodes (Kintec), whose optical constants were measured elsewhere [14]. Our calculation indicates that the transparent efficiency of this ITO layer with 80nm thickness is around 85%, which is insufficient for a high-performance perovskite solar cell. Furthermore, the material quality of the ITO layer is associated with different manufacturing processes. Previous studies have reported the optical properties of the ITO based on ITO-glass (MERCK, Germany) [34], suggesting that this ITO has better transparency and lower absorption than that for the ITO based on Kintec. Thus, we replace the ITO used in our previous calculation with this better ITO to compress the parasitic absorption in the ITO layer.

As shown in Fig 5a, the replacement of the better ITO has a slight influence on the total absorption of the perovskite solar cell with slotted prism based light trapping structure, but the light absorption in the active layer is promoted highly due to the reduced parasitic absorption in the ITO layer. In other words, the light loss caused by reflection and parasitic absorption can be minimized by employing the slotted prism based light trapping and a better ITO layer. $J_G$ of the so-designed perovskite solar cell can be improved to reach 23.92 mA/cm$^2$, about 30% larger than that for the solar cell without light management. In addition, the perovskite solar cell with the better ITO layer and the inverted prism structure also has a remarkable capacity to promote the averaged absorption in the active layer to exceed 90% (see Fig 5b). The utilization of the better ITO only adjusts the internal absorption in different layers of the solar cell, and has a little influence on the total optical benefits induced by the proposed light trapping schemes, so the large serviceable angle for the perovskite solar cell with inverted prism based light trapping structure can still be sustained.

**Electrical performance of the solar cell with proposed light management strategies.** Above studies indicate that the optical properties of the perovskite solar cells can be optimized by invoking the proposed light management schemes. We now explore their electrical properties by implementing an electrical simulation calculation. For a reference, we first implement the simulation on a flat perovskite solar cell without light management. We assume that all absorbed light can be transformed into carriers, and the generation rate (G) of the carriers can be approached by the previous optical calculations. Fig 6a shows G profiles of the flat perovskite solar cell for the normal incident light at wavelength 400nm, 500nm, 600nm and 700nm. At the short wavelength, the light cannot penetrate through the active layer, so the carriers are generated only at the top region of the active layer. Meanwhile, the recombination of the carriers mainly occurs in the region where the carriers are abundant. As the wavelength increases, the light can reach to the bottom Ag layer and be reflected back, which leads to the carriers appearing in the whole region of the active layer. The recombination rate (U) of the carriers in the regions close to the carrier extracting layers is larger than that in the middle region of the active layer. However, more generally, the order of the magnitude of G is larger than that of U for the incident light with all wavelengths. If the loss of extraction of carriers is disregarded, the calculated internal quantum efficiency (IQE) of the flat perovskite solar cell can reach to 100% (see Fig 6b), which is consistent with the previous experiment [14]. Such a perfect IQE is mainly due to the extremely low U of the perovskite solar cell, which is associated with high material quality of the perovskite with long life time and high mobility of the carriers [6,14,35-38]. Optimization of the production process to obtain the material with high quality is certainly the most direct route to maintain the best electrical benefit of the perovskite solar cell.

High-efficient light management schemes proposed in this work can offer best optical benefits for the perovskite solar cell, but the folding regions in the proposed light trapping structures can bring undesirable crystal defects that may increase the recombination of the carriers, and may also degrade the material quality of the perovskite, including life time and mobile of the carriers. To facilitate the investigation how much the optical benefit gained by light management can be transformed into the electrical benefit, the defects in folding regions are regardless to simplify calculations, even if this issue is important for the perovskite solar cell. Here, we suppose that the employment of the light management is independent of the electrical properties of the materials, and IQE=100% for the perovskite solar cell is applicable for all proposed schemes.

By implementing the electrical simulation, the current-voltage (I-V) curve of the perovskite solar cell with different light management schemes can be achieved (see Fig 7). Table 1 shows the basis electrical properties of the proposed perovskite solar cells, including short circuit current density ($J_{sc}$), the open circuit voltage ($V_{oc}$), the filling factor (FF) and the conversion efficiency. The calculated $V_{oc}$=1.02 and the efficiency with 16.13% for the perovskite solar cell deposited on a flat glass is very close to the previous experimental results [14]. By employing slotted and inverted prism based light trapping structures, $J_{sc}$ can be promoted to 21.46 mA/cm$^2$ and 21.01 mA/cm$^2$, respectively. Consequently, the conversion efficiency of the perovskite solar cells with two proposed light trapping structures can achieve 18.89% and 18.47%, respectively. The performance of the perovskite solar cells can be further improved by using a better ITO layer. As a result, for the perovskite solar cell with above two designed light trapping structures and better ITO layers,

$J_{sc}$ can reach to 23.92 mA/cm$^2$ and 23.47 mA/cm$^2$, and the efficiency can achieve to 21.16% and 20.75%, respectively. One may see that the maximal efficiency of the perovskite solar cell with designed light management schemes can exceed the previous reported values, which is 31.2% larger than that for the perovskite solar cell without light management.

The above calculations are implemented when the light is in normal incidence. Here, we examine the incident angle dependence for the perovskite solar cells with different light management schemes. As shown in Fig 8, the maximum efficiency of the perovskite solar cells without light management can be found to approach 16.9% when the incident angle around the Brewster's angle (~50°). For the slotted prism employed perovskite solar cell, the enhancement of the efficiency can be sustained for all oblique angles. However, an obvious decrease of the efficiency appears when the incident angle is between 30° and 65° due to the degradation of the light trapping ability induced by the slotted prism structures. By employing the better ITO layer, the averaged efficiency of the solar cell with slotted prism structure can reach 19.66% for the incident angle less than 80°. Moreover, the efficiency of the so-designed solar cell larger than 20% can be maintained when the incident angle within 22°. To enlarge the serviceable angle of the perovskite solar cell, we also propose another light trapping scheme that is based on the inverted prism structures. In such a scheme, an obvious enhancement of the efficiency of the solar cell can be obtained when the incident angle is less than 60°. It is surprising that the efficiency of the solar cell with the inverted prism structure and the better ITO layer can exceed 20% for all oblique angles from 0° to 50°. If the oblique angle is larger than the Brewster's angle (~50°), the reflection at the front surface of the glass and the side of the inverted prism will increase quickly to degrade the light trapping capacity. Overall, by implementing our light management strategies, the maximum efficiency larger than 21% can be obtained by employing the slotted prism and the better ITO layer, and a larger serviceable angle exceeding 50° can be achieved by employing the inverted prism and the better ITO layer.

## Discussion

To achieve high-performance perovskite solar cells, we proposed high-efficient light management schemes to optimize the optical properties of the cells, which includes the design of light trapping structure to suppress the total light reflection and employs the better ITO layer to reduce the parasitic absorption. By implementing a full field optical and electrical simulation on the designed perovskite solar cells, we discover that the slotted and inverted prism SiO$_2$ structure exhibit better capacities to trap light into the cells. With the properly designed two structures, the total reflection can be compressed to below 5%, and the larger serviceable angle can be achieved for the cell with an inverted prism structure. Based on the proposed light trapping structures, the light absorption in the active layer can be further improved by employing the better ITO layer to reduce the parasitic absorption. The calculated conversion efficiency of the perovskite solar cell with the slotted prism structure and the better ITO layer can approach to an impressive value of 21.16%, being 31.2% larger than that for the solar cell without light management. In addition, another light management strategy that includes the inverted prism structure and the better ITO layer can promote the efficiency of the perovskite solar cell to exceed 20% within a large serviceable angle (~50°). We would like to stress that the proposed structures are feasible to fabricate by laser microlithography. Moreover, the extra cost for constructing proposed structures will not shake the price advantage of the perovskite solar cells. The schemes proposed in this work will provide useful guides for developing high-performance perovskite solar cells.

**Methods**

The optical and electrical properties of the solar cell are studied by employing a full field optical and electrical simulation method that involve a self-consistent calculation of Maxwell, Poisson, and carrier transport equations [39-41]. The optical performance simulation is implemented by solving Maxwell's equations in a Finite Element Method (FEM) software package [39]. Because the prism based light trapping structure is asymmetrical in $x$ and $z$ directions, both the transverse electric (TE) and the transverse magnetic (TM) polarized incident light are considered. The final calculations give the averaged results for TE and TM modes. All of optical calculations are executed under a normal incidence unless specified. The complex optical constants for all layers in proposed perovskite solar cell are taken from previous experimental works [14]. The better ITO layer is adopted from the previous report [34]. By performing the optical simulation, we can obtain the optical absorption in each layer of the solar cell, which is given by:

$$A(\lambda) = \int \frac{\omega \varepsilon^{"} |E(\lambda)|^2}{2} dV, \quad (1)$$

where $E(\vec{r},\lambda)$ is the distribution of the electric field intensity at each single wavelength in each layer, $\varepsilon^{"}$ is the imaginary part of permittivity of the materials, $\omega$ is the angular frequency of the incident light. The optical benefits of the solar cell can be assessed by the density of photo-generated current ($J_G$) given by [42]:

$$J_G = q \int \frac{A(\lambda) P_{am1.5}(\lambda) \lambda}{hc} d\lambda, \quad (2)$$

where q is the charge of an electron, c is the speed of light, h is the Planck constant, $P_{am1.5}(\lambda)$ is the spectral photon flux density in solar spectrum (AM 1.5). By assuming that the absorbed light are all used to excite carriers, the generation profile of the carriers can be described by

$$G(\lambda) = P_{am1.5}(\lambda) \cdot \frac{\varepsilon^{"} |E(\lambda)|^2}{2\hbar}. \quad (3)$$

The electrical performance of the solar cell is simulated by solving Poisson's equation and carriers transport equations in the FEM software package [39]. For simplifying the calculation, only direct and Shockley-Read-Hall (SRH) recombinations are considered. The corresponding coefficients of life time and radiative recombination coefficient are taken from Refs. 6,35,43. The trap energy level is set as $E_t=E_i+0.7\text{eV}$ to fit the $V_{oc}$ value in the previous experiment [14], where $E_i$ is the intrinsic Fermi energy of the $CH_3NH_3PbI_3$. Besides, 6.4 $\Omega\text{cm}^2$ series resistance and 1.6 $k\Omega\text{cm}^2$ shunt resistance are applied to the model for calculating the I-V curve of the perovskite solar cell [44]. Other basis parameters of the perovskite solar cell are taken from previous studies [6,45].

This self-consistent method has been proved to be an effective way to calculate the optical and electrical properties of the solar cells [39]. A proper choice of parameters used in calculations is crucial to realize an accurate and reliable simulation. For a perovskite solar cell, the material parameters are related to the fabrication process. The material parameters used in our calculation are taken from typical experimental and theoretical studies.

## Acknowledgments

The authors are benefitted from useful discussions with Z.C. Wang and Q.R. Zheng. This work is supported in part by the MOST of China (Grant No. 2012CB932900 and No. 2013CB933401), the NSFC (No. 11474279), the Strategic Priority Research Program of the Chinese Academy of Sciences (Grant No. XDB07010100), and the China Postdoctoral Science Foundation (2014M550805). Z.-G. Zhu is supported by Hundred Talents Program of the Chinese Academy of Sciences.

## Author contributions

## Additional information

Supplementary Information accompanies this paper at http://www.nature.com/

**Competing financial interests:** The authors declare no competing financial interests.

**Figures:**

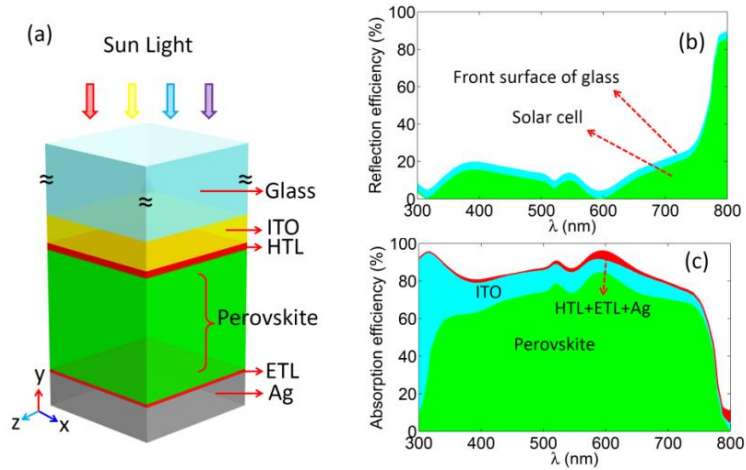

**Figure 1 | Illustration and optical properties of a typical perovskite solar cell.** (**a**) The structure of the perovskite solar cell. (**b**) The reflection efficiency of the solar cell and glass. (**c**) The absorption efficiency of each layer in the solar cell.

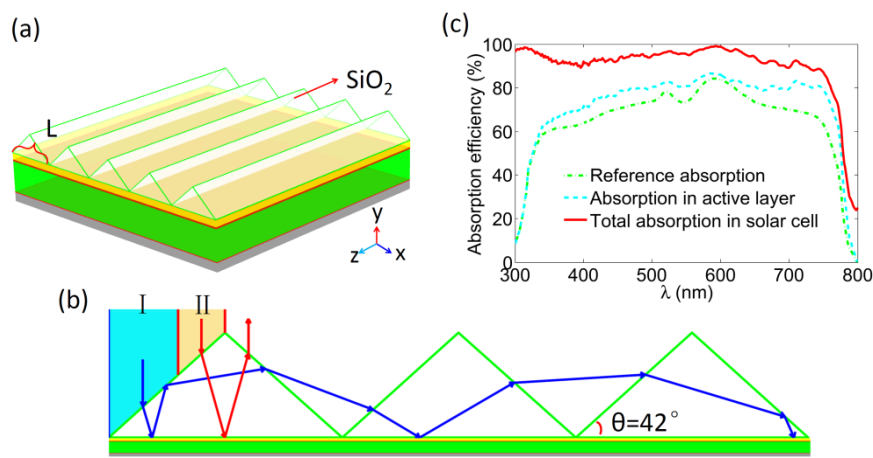

**Figure 2 | Model, light path and optical property of the perovskite solar cell with SiO$_2$ prism light trapping structures.** (**a**) The model of the perovskite solar cell with SiO$_2$ prism structures. (**b**) The schematic of the light ray travel between the prism structures. (**c**) The absorption efficiency of the perovskite solar cell for total and the active layer. The light absorption efficiency in the active layer without light trapping structures is taken as a reference.

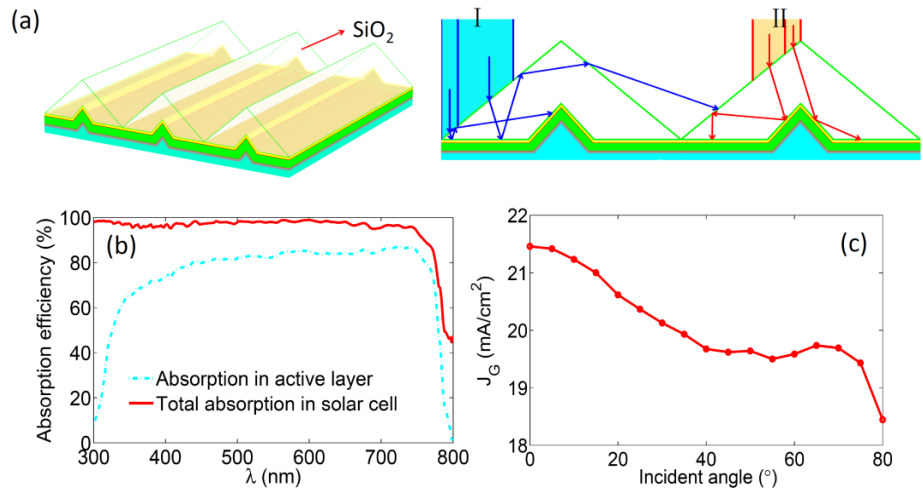

**Figure 3 | Schematic of light trapping induced by slotted SiO$_2$ prism structures and corresponding optical property of the solar cell.** (**a**) The schematic of the perovskite solar cell with slotted SiO$_2$ prism structures and traveling path of the incident light from different regions. (**b**) The absorption efficiency for total and the active layer. (**c**) The photo-generated current (J$_G$) as function of the incident angle.

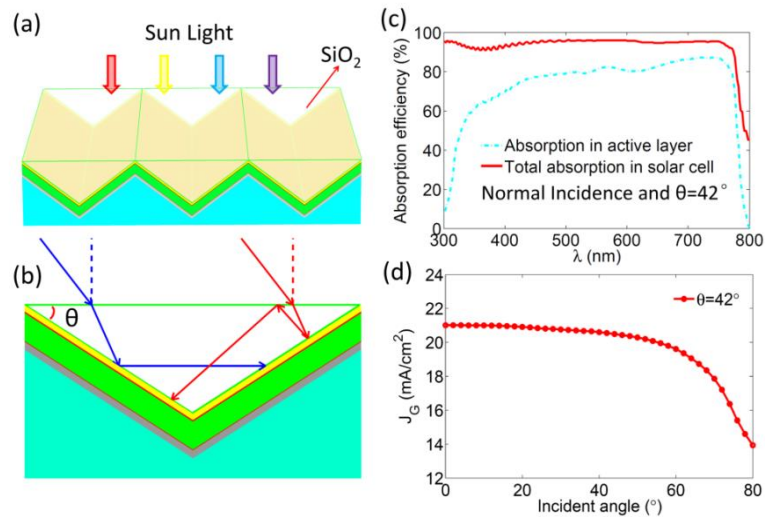

**Figure 4 | Schematic of the light trapping induced by SiO$_2$ inverted prism structures and the corresponding optical property of the perovskite solar cell.** (**a**) The depiction of the SiO$_2$ inverted prism based light trapping structures. (**b**) Sketch of the light trapping principle for the inverted prism structure. (**c**) The absorption efficiency with inverted prism (base angle θ=42°) for total and the active layer. (**d**) J$_G$ with proposed light trapping structure as function of the incident angle.

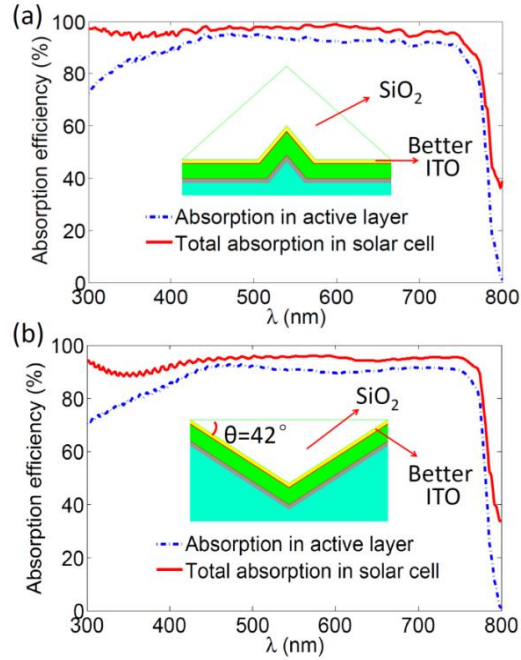

**Figure 5 | Light absorption for the perovskite solar cell with two proposed light trapping structures and the use of the better material quality of the ITO layer.** (**a**) The absorption efficiency of the solar cell with slotted prism and better ITO layer for total and the active layer. (**b**) The absorption efficiency of the solar cell with inverted prism and better ITO layer for total and the active layer.

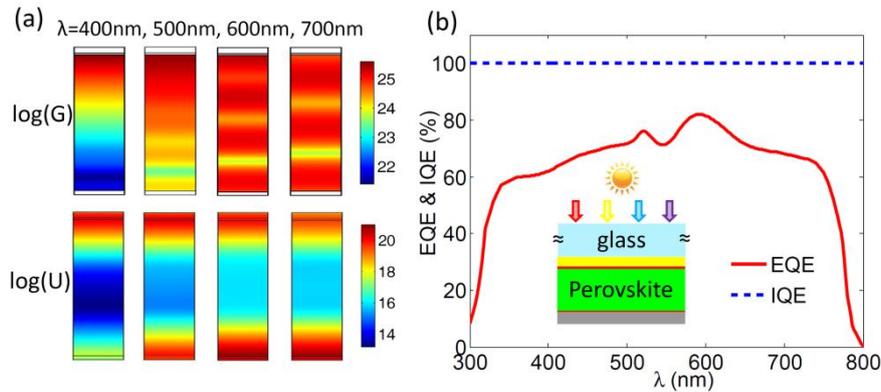

**Figure 6 | Profiles of the generation (G) and recombination (U) rate of the carriers in the flat perovskite solar cell and the corresponding internal and external quantum efficiency (IQE and EQE) of the solar cell.** (**a**) The profiles of log(G) and log(U) of the flat perovskite solar cell without light management. (**b**) The IQE and EQE of the flat perovskite solar cell.

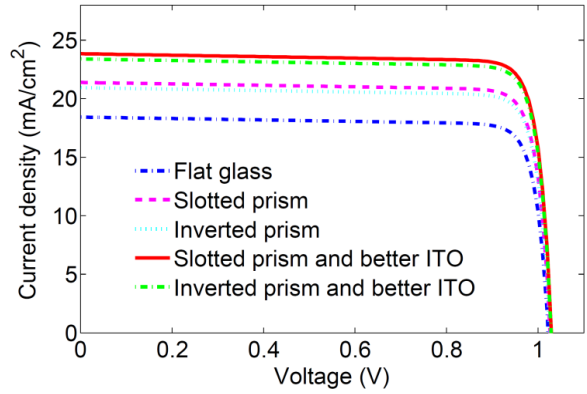

**Figure 7 | Current-voltage (I-V) curve for the perovskite solar cells with different light management schemes.** The current-voltage curve for the flat perovskite solar cell without light management is taken as a reference.

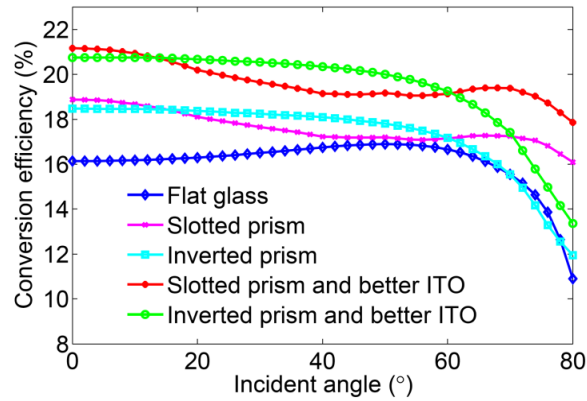

**Figure 8 | Conversion efficiency of the perovskite solar cells with different light management schemes.**

**Table 1** The short circuit current density ($J_{sc}$), the open circuit voltage ($V_{oc}$), the filling factor (FF) and the conversion efficiency of the perovskite solar cells with different light management schemes.

|  | $J_{sc}$(mA/cm$^2$) | $V_{oc}$(V) | FF(%) | Efficiency(%) |
|---|---|---|---|---|
| Flat glass | 18.49 | 1.02 | 85.53 | 16.13 |
| Slotted prism | 21.46 | 1.03 | 85.46 | 18.89 |
| Inverted Prism | 21.01 | 1.03 | 85.35 | 18.47 |
| Slotted prism and better ITO | 23.92 | 1.03 | 85.88 | 21.16 |
| Inverted prism and better ITO | 23.47 | 1.03 | 85.84 | 20.75 |

# Supplementary Information

Highly efficient light management for perovskite solar cells

Dong-Lin Wang[1], Hui-Juan Cui[1], Guo-Jiao Hou[2], Zhen-Gang Zhu[2,1], Qing-Bo Yan[3] & Gang Su[1]*

[1] School of Physics, University of Chinese Academy of Sciences, P. O. Box 4588, Beijing 100049, China.
[2] School of Electronic, Electrical and Communication Engineering, University of Chinese Academy of Sciences, Beijing 100049, China
[3] College of Materials Science and Opto-Electronic Technology, University of Chinese Academy of Sciences, Beijing 100049, China

*Correspondence and requests for materials should be addressed to G.S. (email: gsu@ucas.ac.cn).

**Supplementary Figures**

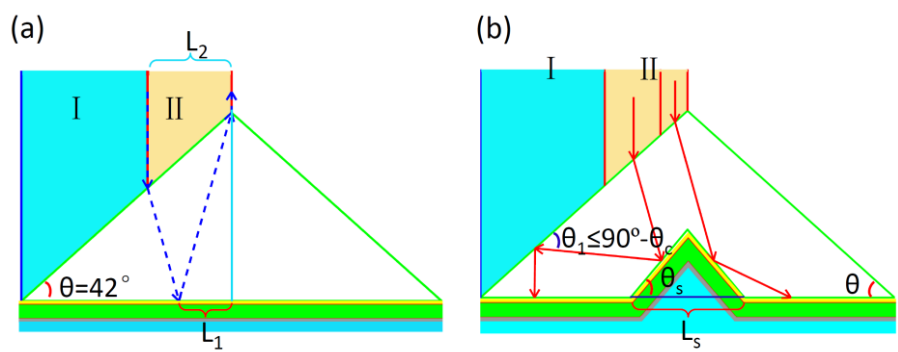

**Figure S1** Schematic of the light trapping induced by prism and slotted prism structures. (a) The light partitions in the perovskite solar cell with prism structure. (b) The representative light rays in the perovskite solar cell with the slotted prism structure.

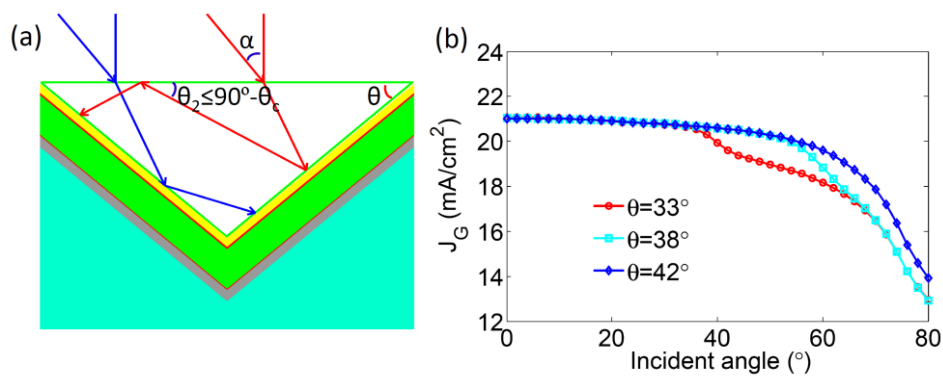

**Figure S2** Schematic of the light trapping induced by the inverted prism structure and the corresponding $J_G$ of the designed perovskite solar cells. (a) The schematic of the light trapping in the perovskite solar cell with the inverted prism structure. (b) $J_G$ of the perovskite solar cell with the inverted prism structure as function of the incident angles.

**Supplementary Note 1:**

**Design of the slotted prism based light trapping structure.** Proper designed prism structure can redirect the light direction to realize an efficient light trapping for the perovskite solar cell. For a prism with the base angle θ=42°, the multiple reflection of the incident light into the active layer can be realized when light comes from I region (see figure S1a). However, light loss is still hard to avoid for the light incident from II region. As shown in figure S1a, the size of the II region is denoted as $L_2$, and the critical size of the solar cell region that corresponds to $L_2$ is marked as $L_1$=L tg(θ) tg(θ-arcsin(sin(θ)/$n_{glass}$))/2. To maximum the light trapping ability of the prism structure, we employ the slotted prism structures to reduce the light loss from II region. As shown in figure S1b, the bottom size of the slotted structure is denoted as $L_s$, and the base angle is denoted as $θ_s$. Here, $L_s$ is set as slightly larger than $2L_1$ to minimize the influence of the light incident from I region. By employing the slotted prism structure, it is easy to realize double rejection of the incident light into the active layer when the light arrives at the right side of the slotted structure. Moreover, the total internal reflection at the prism side can reuse the escaped light from the left side of the slotted structure. To achieve this, a necessary condition that is $θ_1 ≤ π/2-θ_c$ must be satisfied (see figure S1b), where $θ_c$ is the critical angle for the total reflection of light from glass to air. By calculation, we can obtain $θ_1$=π/2-2$θ_s$+2θ-arcsin(sin(θ)/$n_{glass}$), $θ_c$≈41.5°, and also $θ_s$≥49.5°. In this work, L=10μm, $L_s$=2.75μm, and $θ_s$=50° are considered in the simulation of the perovskite solar cell.

**Supplementary Note 2:**

**Design of the inverted prism based light trapping structure.** To enhance the serviceable angle of the perovskite solar cell, we employ the inverted prism structure to realize double rejection of the incident light into the active layer. As shown in figure S2a, we assume the light is oblique incident from the left side of the normal. It is clear that the double rejection of the incident light can be initiatively satisfied when the light arrives at the left side of the inverted prism. To reuse the reflected light that escapes from the right side of the inverted prism, we must ensure that the total reflection at the bottom of the prism is enabled. So, an indispensable condition that needs to be satisfied is $θ_2 ≤ π/2-θ_c$ (see figure S2a). By calculation, we can obtain $θ_2$=π/2-2θ+arcsin(sin(α)/$n_{glass}$), subject to θ>arcsin(sin(α)/$n_{glass}$), where α is the oblique angle of the light. The final solutions for above conditions give α<arcsin($n_{glass}$ sin(2θ-$θ_c$)) for θ≤$θ_c$ and α=90° for θ>$θ_c$. In other words, the serviceable angle for realizing double injection of the light into the active layer can be adjusted by the base angle (θ) of the inverted prism. For example, the serviceable angle around 38° can be obtained by using the inverted prism with θ=33°, and such an angle about 58° can be achieved by the prism with θ=38°. Significantly, the serviceable angle for double rejection of the incident light can be improved to 90° by employing the inverted prism with θ≥42°. By implementing the simulation of the perovskite solar cell, the above principle can be verified by calculating the density of photo-generated current ($J_G$) of the inverted prism at different oblique angles (shown in figure S2b). However, the serviceable angle for the prism with θ=42° only can be sustained within 60° due to the increased reflection occuring when the oblique angle is larger than the Brewster's angle (~50° for the light injecting from air into glass).